# Contra-Analysis for Determining Negligible Effect Size in Scientific Research

**Authors:** Bruce A. Corliss[1,2,*], Yaotian Wang[3], Heman Shakeri[1], Philip E. Bourne[1,2]

**Affiliations:**

[1]School of Data Science, University of Virginia; Charlottesville, Virginia

[2]Department of Biomedical Engineering, University of Virginia; Charlottesville, Virginia

[3]Department of Statistics, University of Pittsburgh; Pittsburgh, Pennsylvania

*Corresponding author. Email: bac7wj@virginia.edu

## Abstract

Scientific experiments study interventions that show evidence of an effect size that is meaningfully large, negligibly small, or inconclusively broad. Previously, we proposed contra-analysis as a decision-making process to help determine which interventions have a meaningfully large effect by using contra plots to compare effect size across broadly related experiments. Here, we extend the use of contra plots to determine which results have evidence of negligible (near-zero) effect size. Determining if an effect size is negligible is important for eliminating alternative scientific explanations and identifying approximate independence between an intervention and the variable measured. We illustrate that contra plots can score negligible effect size across studies, inform the selection of a threshold for negligible effect based on broadly related results, and determine which results have evidence of negligible effect with a hypothesis test. No other data visualization can carry out all three of these tasks for analyzing negligible effect size. We demonstrate this analysis technique on real data from biomedical research. This new application of contra plots can differentiate statistically insignificant results with high strength (narrow and near-zero interval estimate of effect size) from those with low strength (broad interval estimate of effect size). Such a designation could help resolve the File Drawer problem in science, where statistically insignificant results are underreported because their interpretation is ambiguous and nonstandard. With our proposed procedure, results designated with negligible effect will be considered strong and publishable evidence of near-zero effect size.

# Introduction

Two-sample p-values from null hypothesis significance tests remain the gold standard for the analysis and reporting of scientific results despite calls to discontinue or de-emphasize their use (*1*, *2*). P-values can separate results based on statistical significance, where positive results (statistically significant) demonstrate sufficient evidence of a non-zero effect size, while null results (statistically insignificant) have an effect size that plausibly might be zero. Yet p-values do not yield a practical interpretation of effect size (*3*), and there is no standard method to do so. A practical interpretation would determine if the observed effect size were large enough to be meaningful in the real world, small enough to be negligible, or neither (Fig 1A). Previously, we proposed contra-analysis as a method of analyzing results that determines if the effect size from controlled experiments has evidence of being practically meaningful. We developed contra plots to visualize effect size across various interventions from broadly related experiments. We define broadly related experiments as those that measure the same phenomenon across experiment designs with different treatments, timepoints, experiment models, and experiment conditions. Contra plots can be used to score and rank effect size across experiments, aid researchers in determining a threshold for meaningful effect, and perform a hypothesis test to determine which interventions have evidence of meaningful effect size.

While this process is helpful in providing a structured method to analyze results for meaningful effect, contra-analysis does not help determine if an effect size is small enough to be considered negligible (approximately zero). Determining if an effect size is negligible is important in scientific research for eliminating alternative scientific hypothesis, asserting approximate equivalence between study groups, and declaring approximate independence between the intervention and the measured variable (Fig 1B-E). Such an assertion could strengthen a result with meaningful effect size by eliminating alternative explanations (Fig 1F) or determining if unwanted side-effects associated with an intervention are acceptably negligible for real world use (Fig 1G). These tasks are critical for establishing causal and non-causal relationships between variables in scientific studies (*4*).

Statisticians have developed several statistics to conclude negligible effect, including the two-one sided t-test equivalence p-value (*5*), bayes factor (*6*), and second-generation p-value (*7*) to fulfill this need. Yet all these statistics do not inform the actual selection of the threshold for negligible effect. Such a threshold would represent the maximum upper limit for the relative difference in means that would still be considered a negligible effect size. Determining the value of the threshold for negligible effect should consider the precedence established by previous research. Considering precedence requires visualizing and comparing effect size across broadly related experiments. All reported results, whether their effect size is meaningful, negligible, or inconclusive, can potentially influence the selection of a maximum threshold for negligible effect.

We outline a secondary use of contra-analysis and contra plots to analyze negligible effect size. We propose a decision-making procedure that uses contra plots to inform the selection of a threshold for negligible effect size, determine which results have negligible effect with a hypothesis test, and score results based on how negligible the effect size is. We demonstrate the process on real biomedical research data. With our proposed procedure, results designated with negligible effect may be considered strong and publishable evidence of near-zero effect size.

## Contra-analysis Procedure for Negligible Effect

We propose contra plots as a data visualization that can be used to determine if a result demonstrates evidence of a negligible effect size. Determining if an effect size is negligible is not just useful for discussing results in a publication, but also for informing decisions throughout the research life cycle, from ideation and pilot data collection to publishing and translation (Fig 2). Contra plots summarize effect size from broadly related experiment results that measure the same phenomenon. This data visualization is designed as a central step in the process of contra-analysis, where the practical interpretation of broadly related experiment results is continuously assessed relative to one another. In contra-analysis, scientists carry out a continuous decision-making cycle that compares the effect size of various interventions on a common measured phenomenon (Fig 2A). A central step in this cycle is the use of contra plots.

A contra plot is a data visualization composed of a plot and table juxtaposed horizontally (*8*), similar in design to forest plots used for meta-analysis (*9*). The plot region visualizes the credible interval estimates of effect size from each experiment and the table region contains metadata about the experiments the estimates are derived from (Fig 2B). The interval estimates have a multiple comparisons correction that sets the credible level of each study to 0.95. Following the practice of previous studies (*10–13*), we use a Bonferroni correction for credible intervals. We focus on experiments that measure a continuous variable that have an experiment group and control group. We use the relative difference in means to measure effect size for this case (difference between experiment and control sample means divided by control sample mean). The interval estimates are horizontally displayed and aligned by row with metadata from the experiment they belong to. The content of the metadata is flexible and should be tailored to maximize interpretability. The metadata would at least include a description of the intervention used and the originating study, but could also include information about the species, control group, experiment model, timepoints, dosages, disease model, and population studied. We strongly recommend providing supplemental metadata in a separate table so that scientists can look-up additional information if necessary (see examples in Tables S1, S2).

With our previously proposed contra plot procedure for identifying meaningful effect, we included a statistic in the metadata table labeled as Ls% (denoting the "least percent" within the interval estimate) that aids with scoring, ranking, and performing hypothesis tests for meaningful effect (*8*). Here, we propose a second statistic, listed in the Ms% column, that will aid with scoring, ranking, and performing hypothesis tests for negligible effect size. In this case, Ms% denotes the upper bound of the credible interval of the unsigned relative difference in means (essentially an estimate of its largest plausible absolute value, conveyed as a percent). We use the absolute value because we evaluate negligible effect by its distance to zero- the sign of effect is irrelevant for this case. For example, a Ms% value of 10% at a 95% credible level would mean there is a 0.95 probability that the unsigned relative difference in population means is between -10% and +10% (see Methods and Supplementary Methods section for more details and a mathematical explanation, where Ms% is referred to as $r\delta_M$ for the name of the statistic). A smaller Ms% value is considered more negligible.

We provide a brief overview of contra-analysis for analyzing negligible effect size. A more detailed explanation of various parts of this process is found in the Methods section (see Bayesian Summary of Difference in Means, Scoring and Ranking Negligible Effect Size, and Hypothesis Testing). Several steps are taken directly from our procedure for contra-analysis for meaningful effect since there is overlap between the two processes (*8*).

1. **Compile related results**: to start this process, scientists compile results that measure the same phenomenon. These results should be from studies that have passed peer review. The scientist will record summary statistics of the data into a CSV file, including the mean, standard deviation, and sample size of the control group and experiment group (see supplementary files 1 and 2 for examples). These can be directly copied from the text, obtained from correspondence with the authors, or estimated from the plots that visualize the data. The scientist will also record various information about the study for the metadata table in the contra plot and any supporting supplementary tables. Each study is assigned a unique identifier number (displayed on left side of plot) so scientists can look-up additional information in the supplementary metadata table (see examples in Tables S1, S2).

2. **Produce contra plot:** once related results are collected, scientists will run the contra plot function to generate the data visualization (Fig 2B).

3. **Score interventions by negligible effect size**: the studies displayed in contra plots are scored by negligible effect, which is listed in the Ms% column (see Fig 4 for example). For illustrative purposes, this score can *approximately* be visualized by the magnitude of the outermost credible interval bound from the origin (marked by blue lines, Fig 3C). This is an approximation because the interval estimates displayed in contra plots are for the signed effect size rather than the unsigned effect size used to calculate Ms%. The studies that have an interval estimate that includes zero are sorted in such a way that the most negligible effect sizes are located at the vertical center of the contra plot and become less negligible above and below that position. With this sorting, all studies that are below the threshold will be found at the vertical center of the contra plot.

4. **Determine threshold for negligible effect**: the scientist determines the threshold for a negligible effect size. This threshold is the largest value for the relative difference in means that would still be considered negligible (Fig 3D). While scientists can specify whatever threshold they wish, they need an evidence-based justification for its value. Contra plots facilitate this discussion with a straightforward visualization and scoring of negligible effect size. An appropriate threshold would ideally pass results that are widely thought to provide high strength evidence of negligible effect. A threshold must also balance those results against the requirements of potential real-world considerations (e.g., acceptable effect size for unwanted side effects of a treatment). Threshold choice could also give special consideration for studies measured from the final model system (e.g., the human body in biomedical sciences, with results reported from clinical records or from clinical trials). Justification for the chosen threshold should be explicitly discussed in the text describing the result.

5. **Identify interventions with negligible effect**: scientists perform a hypothesis test to determine which results have negligible effect size based on the specified threshold (Fig 3E). The results with a Ms% value that is less than the specified threshold are designated as having negligible effect size, while results that are greater than the threshold are not negligible.

6. **Assert negligible effect**: scientists have several options to describe results with negligible effect size. Scientist can assert approximate independence between the intervention and the measured variable or approximate equivalence between study groups. They can also conclude that approximately no effect or change is associated between the intervention and the measured variable.

7. **Obtain new results**: the new assertion adds to the web of scientific knowledge and influences the next research inquiry pursued and future experiments performed.

8. **Recompilation**: the new experimental results are incorporated into an updated version of the same contra plot, and the decision-making process of contra-analysis will repeat.

Contra plots could be shared as research output in review papers or presented as a dynamic web application that continuously adds new studies to the plot as they are published. For primary research papers, contra plots can be used to justify the claims for the impact and practical significance of reported effect sizes (*3*).

## Results

We include two case studies developed from our previous work on contra-analysis to illustrate how to identify results with negligible effect size. We compiled a collection of results from different studies with different experiment designs that measure the same phenomenon. In this case, total plasma cholesterol and plaque size are the measured phenomena used to illustrate the use of contra plots for analyzing negligible effect.

**Case Study 1: Total Plasma Cholesterol**

Total plasma cholesterol (TPC) is an important clinical indicator for evaluating patients with atherosclerosis. While lowering total plasma cholesterol is therapeutic in most cases (depending on the composition of the cholesterol (*14*)), many interventions treat atherosclerosis through other means. It is important to determine whether an intervention has a negligible effect on total plasma cholesterol to help elucidate its underlying mechanisms. Plasma cholesterol levels vary from 60-3000 mg/dL across animal models used to research atherosclerosis and are reported in units of mmol/L as well (Table S1). This large variation in the measurement values makes it necessary to evaluate negligible effect on a relative scale. Determining the threshold for negligible effect size should be influenced by what is considered a meaningful effect size. Since treatments that reduce TPC have been used in the clinic, determining a threshold for effect size should consider current gold standard treatments and previous research studies. Statins are considered a gold standard treatment and clinical results indicate a 30% - 60% reduction in cholesterol (*15*) (only the LDL-C subtype of cholesterol was measured for these results, but for the purpose of this case study, we are disregarding the complexities associated with subtypes). An initial maximum threshold of 30% for effect size would be below most of the reported effect sizes for the gold standard and still designate some studies with negligible effect size (*16*). Referring to the contra plot, the threshold of 30% also provides some degree of separation between null results and those with a nonzero effect size (Fig 3A).

Inspecting the contra plot will readily reveal the studies that have Ms% < 30% (studies 22, 31, 1, 2) and are designated as having evidence of negligible effect size for reducing TPC. Using Ms% as a scoring metric, we see there is a 10-fold spread of negligible effect score within the collection of interventions that lack statistical significance (roughly corresponding to results with Ls% value of 0, although a different statistical analysis was used in each study from our use of credible intervals). This highlights the need to analyze the practical interpretation of effect size with statistically insignificant results because they can offer different levels of evidence for negligible effect that span orders of magnitude. Additionally, there is a two-fold range of scores within the results designated as having negligible effect. This highlights that not all interventions with negligible effect are equal. While all the cited publications in the table with null results correctly stated that no difference was observed between the control and experiment group, most results were still used indirectly as evidence of negligible effect size either in the text or as a secondary negative control. With the specific threshold, scientists could not use results with Ms% ≥ 30% as negligible effect. Instead, they could choose to present the data with an ambiguous interpretation or collect additional samples in an attempt to clarify the interpretation of effect size.

Critically, this analysis can be repeated by multiple scientists using any threshold from a single static contra-plot. The interval estimates in contra plots are invariant to the value of the threshold. This allows scientists to form their own opinion for an appropriate threshold and facilitates discussion and consensus between scientists. More importantly, a single contra plot can be applied to multiple related contexts. For example, a contra plot could aid in determining different thresholds for negligible effect size for unwanted side effects of long-term early-stage atherosclerosis treatments (with lower tolerance for unwanted side-effects) versus acute late-stage atherosclerosis treatments (with higher tolerance for unwanted side-effects).

**Case Study 2: Plaque Size**

As a second example, a similar case study examined interventions that reduced, increased, or had no discernable effect on plaque area as a treatment for atherosclerosis (Fig. 4A). Similar to measuring total cholesterol, plaque size is measured across units that span orders of magnitude (Table S2). Using the same general approach as the TPC example, a maximum threshold of 35% change would reveal that studies 1, 2, 3, and 4 have a negligible effect size. For these cases, scientists could describe those results as having evidence of negligible effect.

## Discussion

We propose a process to analyze and evaluate evidence of negligible effect size for results obtained from controlled experiments. Our process relies on contra plots, a data visualization that displays interval estimates of effect size for a common measured phenomenon across broadly related experiments. To facilitate this analysis, we developed a statistic (labeled as Ms%) that estimates the upper bound of the unsigned relative difference in means. Using Ms% and contra-plots, scientists can compare effect size across different experiment designs, score results based on their negligible effect, determine a threshold informed by the precedence of related results, and perform a hypothesis test to identify which results have negligible effect. Determining negligible effect is critical in scientific research for ruling out alternative hypotheses, establishing approximate equivalence between study groups, and identifying approximate independence between an intervention and a measured phenomenon. We illustrate how to carry out these tasks using real biomedical research data.

A result's effect size has no practical meaning when evaluated in isolation- it must be considered within the context of potential applications and related results. A contra plot can summarize all related results in a single visualization so scientists can quickly gain a perspective of the precedence established by previous research. For determining negligible effect size, a contra plot will aid in revealing how negligible previous results are and give a perspective for what effect sizes are practically meaningful, both of which are important for determining a threshold for negligible effect. No other data visualization and statistic allows comparing negligible effect size across broadly related experiments. Furthermore, contra-plots allow scientists to perform an on-the-fly hypothesis test against any threshold value for negligible effect. The chosen threshold may differ for the same collection of results when evaluated within different research subfields (with unique applications), by scientists with different perspectives, or across time as expectations become more stringent as technology improves with measurement techniques.

Statisticians have developed several statistics to analyze negligible effect, including the two-one sided t-test equivalence p-value (*5*), bayes factor (*6*), and second-generation p-value (*7*). Yet there is no clear way these statistics can visualize and summarize effect size across broadly related experiments, inform the choice of a threshold, and allow for on-the-fly reanalysis of results against any threshold value. Our approach supports these tasks because it summarizes the sample data rather than the extent that the sample data supports a specific hypothesis. We believe summarizing sample data with a specific hypothesis offers an overly narrow summary of effect size that limits its practical interpretation. Furthermore, these statistics are not designed specifically for use with relative units, and consequently cannot easily summarize and compare effect size across broadly related experiments.

Contra-analysis, contra plots, and our statistics (Ls% and Ms%) are presented as a decision-making procedure for comparing effect size between different experiment designs rather than a formal statistical analysis applied to a single experiment. Yet our procedure could potentially be used as a statistical process within a single study where the authors designate and justify their thresholds for meaningful and negligible effect size and perform hypothesis tests for both when reporting results. To do so, the statistical properties of Ls% and Ms% must be further characterized and formalized with further research (*17*, *18*). Contra plots currently use a Bayesian approach with credible intervals to estimate effect size. While there is no known method to calculate the confidence interval of the unsigned relative difference in means, further research may yield a method to do so and provide a frequentist version of the contra plot.

This new application of contra plots could also provide a solution to the File Drawer problem (*19*), where statistically insignificant results are underreported because their interpretation is ambiguous. A primary reason for this phenomenon is that a scientist cannot conclude much from a null result other than no change was observed- they cannot conclude that no change existed. Contra plots can be used to identify which statistically insignificant results have evidence of negligible effect size by differentiating high strength results with a near-zero effect size range from low strength results with a broad range of plausible effect sizes. Providing a clear and standard interpretation of statistically insignificant results may encourage their publication, prevent scientists from needlessly repeating null experiments that have already been done, and increase the overall rigor of scientific research (*19*).

## Methods

### Code Availability

All code used to generate all figures is written in R and available at: https://github.com/bac7wj/contra. Code was executed using RStudio 2022.07.0+548 "Spotted Wakerobin" Release for Windows (Windows NT 10.0; Win64; x64).

### Bayesian Summary of Difference in Means

We define the equations and assumptions used to calculate the raw and relative difference in means. Let $X_1, ..., X_m$ be an i.i.d. sample from a control group with a distribution Normal($\mu_X, \sigma_X^2$), and $Y_1, ..., Y_n$ be an i.i.d. sample from an experiment group with a distribution Normal($\mu_Y, \sigma_Y^2$). Both samples are independent from one another, and we conservatively assume unequal variance, i.e., $\sigma_X^2 \neq \sigma_Y^2$ (the Behrens-Fisher problem (*20*) for the means of normal distributions).

We analyze data in a Bayesian manner using minimal assumptions and therefore use a noninformative prior, specified as

$$p(\mu_X, \mu_Y, \sigma_X^2, \sigma_Y^2) \propto (\sigma_X^2)^{-1} (\sigma_Y^2)^{-1}. \tag{1}$$

The model has a closed-form posterior distribution. We define the sample means $\bar{x}$ and $\bar{y}$, along with variances $s_X^2$ and $s_Y^2$ for the control group and experiment group. Specifically, the population means,

conditional on the variance parameters and the data, follow normal distributions:

$$\mu_X | \sigma_X^2, x_{1:m} \sim \text{Normal}\left(\bar{x}, \frac{\sigma_X^2}{m}\right) \text{ and} \tag{2}$$

$$\mu_Y | \sigma_Y^2, y_{1:n} \sim \text{Normal}\left(\bar{y}, \frac{\sigma_Y^2}{n}\right). \tag{3}$$

The population variances each independently follow an inverse gamma distribution (InvGamma):

$$\sigma_X^2 | x_{1:m} \sim \text{InvGamma}\left(\frac{m-1}{2}, \frac{(m-1)s_X^2}{2}\right) \text{ and} \tag{4}$$

$$\sigma_Y^2 | y_{1:n} \sim \text{InvGamma}\left(\frac{n-1}{2}, \frac{(n-1)s_Y^2}{2}\right). \tag{5}$$

We exclude the use of prior information in this analysis because we wish to summarize the data alone and not be influenced by the beliefs of the scientist reporting the data (specifically, the strength of the prior used can considerably influence the outputs of a Bayesian statistical analysis (*21*, *22*)).

We define the unsigned relative difference in means ($|r\mu_{DM}|$) as:

$$|r\mu_{DM}| = \left|\frac{\mu_Y - \mu_X}{\mu_X}\right|. \tag{6}$$

This variable is estimated to score results from broadly related experiments based on negligible effect size.

**Scoring and Ranking Negligible Effect Size**

Interval estimates of relative effect size and the Ls% statistic within contra plots were calculated with credible intervals of the relative difference in means as done previously (*8*). For calculating the Ms% statistic used for scoring negligible effect size, we used a similar strategy by calculating the upper bound of the credible interval of the unsigned relative difference in means. Just as with the Ls% statistic, Ms% is calculated with repeated Monte Carlo samples of the posterior distribution (see Supplemental Methods: Interval Estimation for more details, where Ms% is referred to as rδ$_M$ for the actual name of the statistic). Our statistic corresponds to a credible interval centered about zero that is bound by the positive and negative value of Ms%. This statistic can be interpreted the same as the bound of any other credible interval. For example, a Ms% value of 10% at the 0.95 credible level would mean that the relative difference in means has a 0.95 probability of being between -10% and +10%. Credible intervals were used because confidence intervals cannot be calculated for the unsigned relative difference in means because no pivotal quantity exists for this case.

**Hypothesis Testing**

Once a threshold is specified, a scientist can identify which interventions have evidence of a negligible threshold by performing an interval-based hypothesis test. Such a test can be done on-the-fly by checking if Ms% for each study is less than the threshold (see Supplemental Methods: Hypothesis Testing for details). As mentioned above, the Ms% statistic represents a zero-centered credible interval bound by [-Ms%, +Ms%]. This test would check that all the plausible values for the unsigned effect size (those contained within the zero-centered credible interval) are smaller than the threshold specified. Although interval estimation is typically used for parameter estimation, hypothesis testing can be performed with credible intervals. The procedure checks if the threshold value is within the bounds of the interval (*23*).

An advantage of using an interval estimation approach for hypothesis testing is that scientists can perform the test using any threshold without requiring recalculation of a statistic or interval. This property of interval estimates allows the same contra plot to be used for hypothesis testing regardless of the threshold value specified. Therefore, scientists with different perspectives or studying different applications can use the same contra plot to perform this analysis.

# Figures

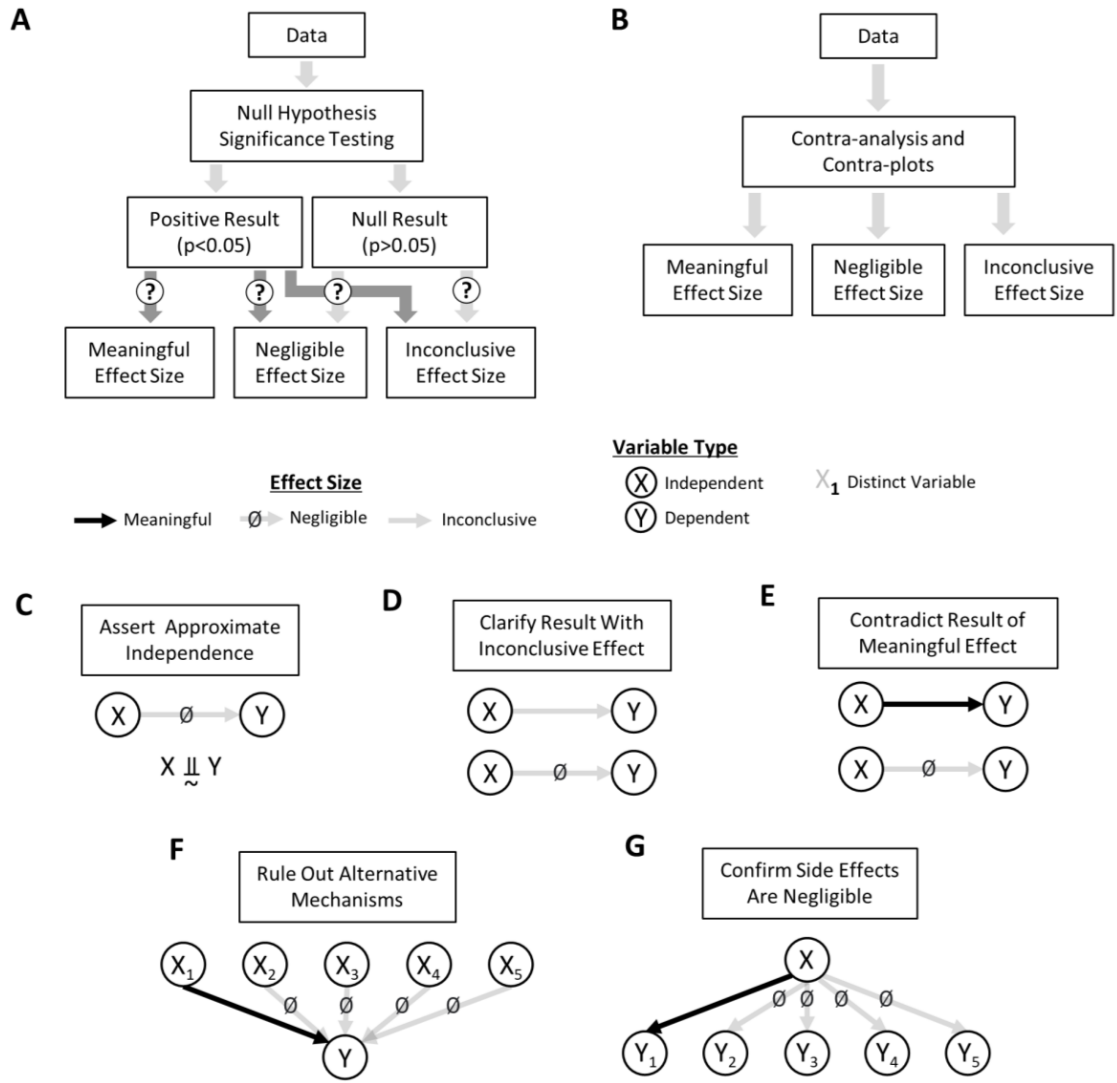

**Figure 1: Illustration of contra-analysis and the uses of determining negligible effect size**. (**A**) Flowchart for analyzing results with NHST methods, highlighting the ambiguity and lack of standards for classifying the practical significance of effect size. (**B**) Flow chart for contra-analysis illustrating a more direct and structured method to classify effect size. (**C-G**) Diagrams illustrating possible uses for determining negligible effect size in controlled experiments, where X is an independent variable encoding the presence and absence of an intervention (i.e. experiment study group and control study group), and Y is a dependent continuous variable (numeric subscripts for both denote different variables that are unrelated). (**C**) If the presence of X has a negligible effect size with Y, then X is approximately

independent of Y (denoted with an independence symbol over a tilde), and the study groups with and without X are approximately equivalent. (**D**) If a previous result indicates an inconclusive effect size (not negligible) with the presence of X with Y, then a repeated experiment with negligible effect size would clarify the relationship. (**E**) If a previous result indicates a meaningful effect size with the presence of X with Y, then a repeated experiment with negligible effect size would strongly contradict the previous result. (**F**) If there are several possible mechanisms (tested with interventions $X_{1-5}$) responsible for changing Y and only the presence of $X_1$ has meaningful effect, then asserting negligible effect with $X_{2-5}$ with Y would strengthen the finding by eliminating alternative explanations. (**G**) If $Y_1$ is a variable that measures a desired effect while $Y_{2-5}$ describe undesirable effects, a scientist could confirm that the intervention has negligible side effects with $Y_{2-5}$.

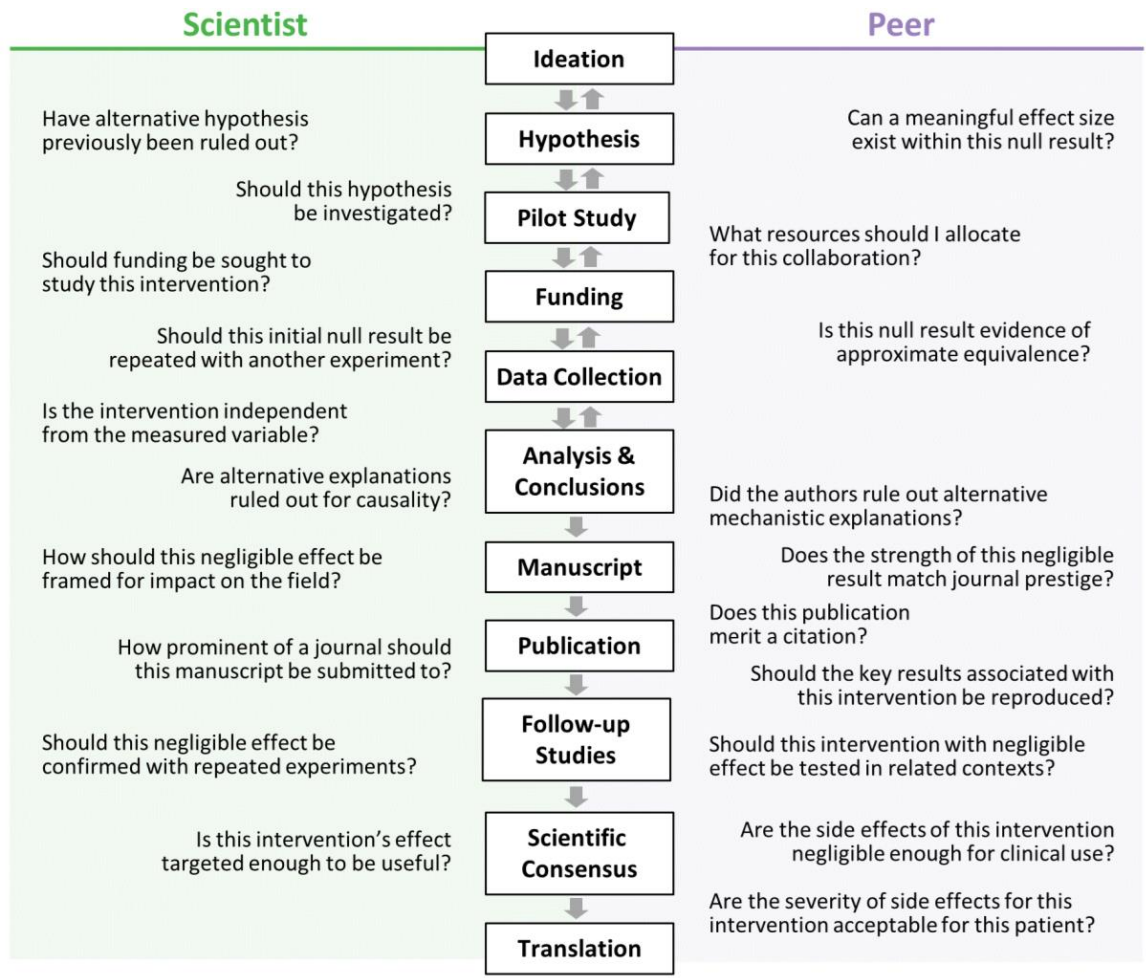

**Figure 2: Contra-analysis can inform critical decision-making steps in the scientific research lifecycle**. Flow chart of stages of the process of scientific research, from ideation to translation. For both the scientist pursuing the research (left column) and their peers (right column), a series of questions are posed that represent key decision-making steps that are informed by comparing negligible effect size between different interventions on the same measured phenomenon. Contra-analysis could be used to inform the decision-making to answer these questions (in conjunction with a plethora of practical, social, and philosophical considerations that must also be considered).

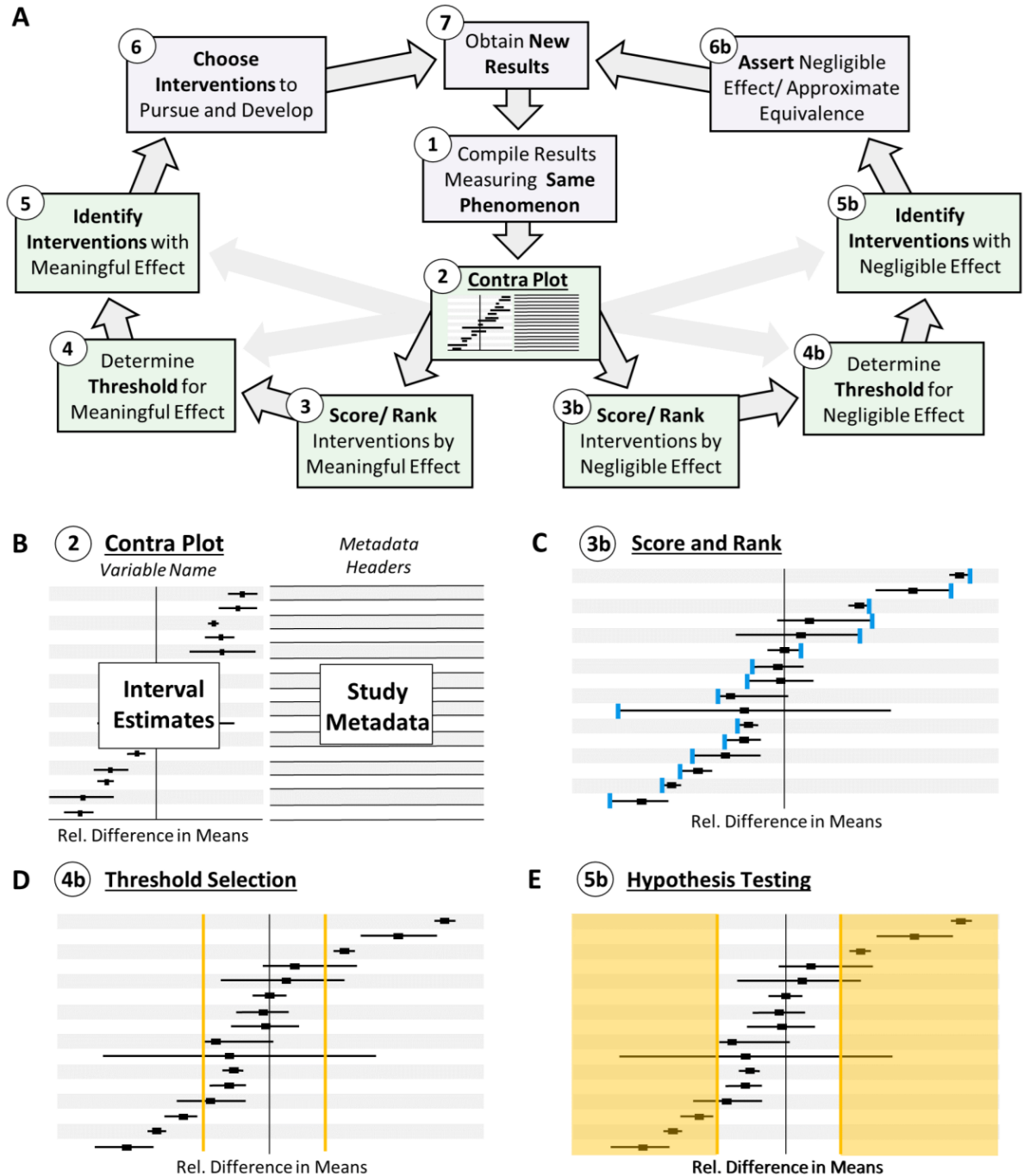

**Figure 3: Contra-analysis process flowchart and procedure for analyzing negligible effect with contra plots**. (**A**) Flow chart for contra-analysis, where scientists compare effect size across dissimilar experiments that measure the same phenomenon and classify which interventions show evidence of meaningful or negligible effect size. (**B**) Basic structure of a contra plot that includes interval estimates of effect size paired with a table of metadata describing the originating study for each result. (**C**)

Visualization of negligible effect score approximated as the unsigned outermost interval bound from the origin (blue line, this is an approximation because the actual score is calculated from the unsigned effect size while the interval estimates are for the signed effect size). (**D**) Visualization of threshold selection. (**E**) Illustration of hypothesis testing for negligible effect size with contra plots, where results with intervals entirely within the negligible region are considered negligible (between gold lines).

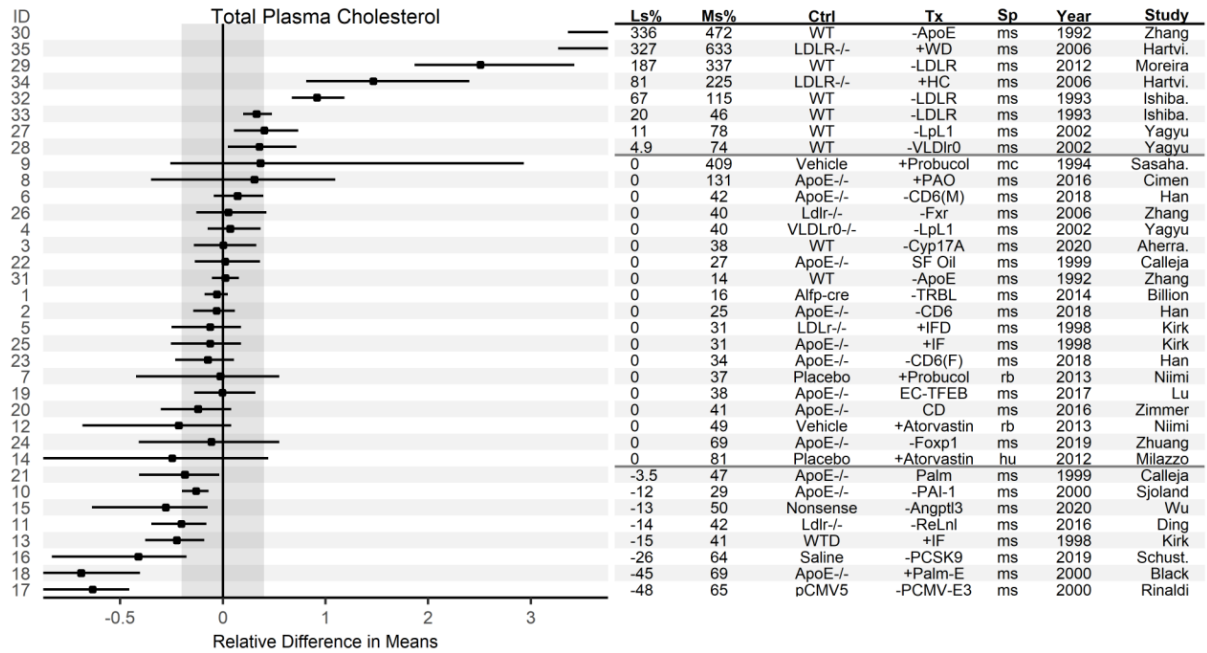

**Figure 4: Visualizing effect size across different experiments measuring total plasma cholesterol.** Contra plot of interventions that either reduced, increased, or had no discernable effect on total plasma cholesterol (for illustrative purposes, grey region centered at origin is maximum threshold of 30% for negligible effect). Abbreviations: Ls%, closest interval bound to zero (the "least" value of the interval), expressed as a percentage; Ms%, furthest interval bound to zero (the "most" value of the interval), expressed as a percentage (calculated from unsigned effect size rather than the signed effect size used for visualized interval estimates); Sp, species for experiment model; Ctrl, control group label; Tx, treatment label for experiment group. Intervals are 95% credible intervals of the relative difference in means, Bonferroni corrections are applied to design of each individual study (See STable 1) with no additional correction controlling for the number of studies included in the contra plot.

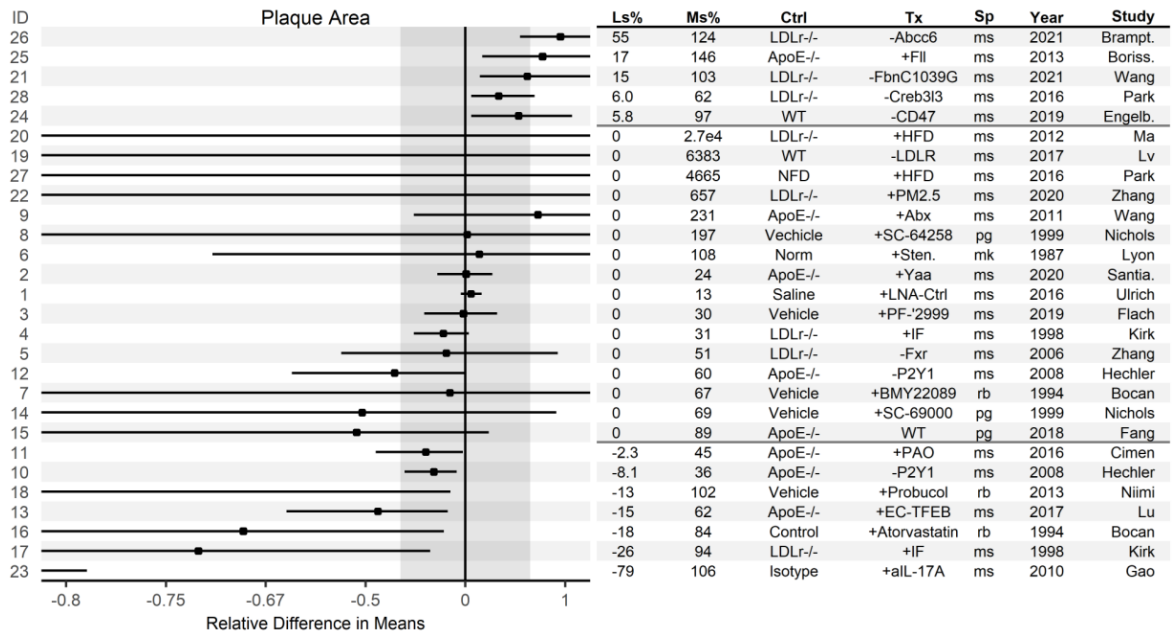

**Figure 5: Visualizing effect size across different experiments measuring total plasma cholesterol.** Contra plot of interventions that either reduced, increased, or had no discernable effect on plaque area (for illustrative purposes, grey region centered at origin is maximum threshold of 35% for negligible effect). See Fig. 3 for abbreviations. Intervals are 95% credible intervals of the relative difference in means, Bonferroni corrections are applied to design of each individual study (See STable 2) with no additional correction controlling for the number of studies included in the contra plot.

# Supplementary Materials for

## Contra-Analysis for Determining Negligible Effect Size in Scientific Research

**Authors:** Bruce A. Corliss[1,2,*], Yaotian Wang[3], Heman Shakeri[1], Philip E. Bourne[1,2]

**Affiliations:**

[1]School of Data Science, University of Virginia; Charlottesville, Virginia

[2]Department of Biomedical Engineering, University of Virginia; Charlottesville, Virginia

[3]Department of Statistics, University of Pittsburgh; Pittsburgh, Pennsylvania

*Corresponding author. Email: bac7wj@virginia.edu

**This PDF file includes:**

Supplementary Materials and Methods

Tables S1 to S2

Supplementary References

# Supplementary Methods and Materials

## Interval Estimation

We wish to score effect size estimates based on how negligible (near-zero) they are. We score near-zero effect size by its distance from zero. The direction of effect is disregarded since a positive and negative effect size of the same magnitude are equally negligible. For measuring a continuous variable, we estimate the unsigned relative difference in means ($|r\mu_{DM}|$), where

$$|r\mu_{DM}| = \left|\frac{\mu_Y - \mu_X}{\mu_X}\right|. \tag{S7}$$

We estimate this quantity by calculating the credible interval for $|r\mu_{DM}|$. While there is no closed-form posterior distribution for $|r\mu_{DM}|$, we can estimate its upper quantile using Monte Carlo simulations of the same posterior distribution derived from the prior and likelihood introduced in Eq. (4) – (5). We exploit the fact that $-c < r\mu_{DM} < c$ if and only if $|r\mu_{DM}| < c$, and we define $F_{relative}(x)$ as the empirical cumulative distribution function of the signed relative difference in means approximated with Monte Carlo simulations from the posterior:

$$F_{relative}(x) = K^{-1} \sum_{i=1}^{K} \mathbb{I}\left(\frac{\mu_Y^i - \mu_X^i}{\mu_X^i} \leq x\right). \tag{8}$$

We could in theory calculate a lower ($\hat{c}_{lo}$) and upper ($\hat{c}_{hi}$) quantile of the estimate for $|r\mu_{DM}|$, which would enclose a set of plausible values for $|r\mu_{DM}|$. Instead, we score negligible effect size based on the least negligible value contained within the interval. The least negligible value is the furthest value from zero. This value is the upper quantile of the posterior of $|r\mu_{DM}|$, calculated by solving for $\hat{c}_{hi}$ such that

$$F_{relative}(\hat{c}_{hi}) - F_{relative}(-\hat{c}_{hi}) = 1 - \alpha_{DM}. \tag{9}$$

Because this interval is centered at 0 (i.e. $(-\hat{c}_{hi}, \hat{c}_{hi})$), we report only the upper tail. We name this statistic the relative most difference in means ($r\delta_M$). Specifically, if $Q_{relative}(p)$ is the quantile function of the posterior $F_{relative}(|x|)$, then $r\delta_M$ satisfies

$$r\delta_M = Q_{relative}(1 - \alpha_{DM}). \tag{10}$$

The r$\delta_M$ is associated with a percentage (1 - $\alpha_{DM}$) to denote the credible level (i.e., a 95% r$\delta_M$ is calculated with a credible level of 0.95). Colloquially, the value of r$\delta_M$ represents the largest unsigned percent difference between the population means of the experiment group and control group supported by the data. Results with lower values of r$\delta_M$ are considered more negligible. This statistic is labeled as Ms% in contra plots for ease of interpretation, since the "most %" is an intuitive label for conservatively evaluating how near-zero an effect size estimate is.

**Hypothesis Testing**

To determine if a result has a negligible effect size, scientists can perform a hypothesis test by testing if |r$\mu_{DM}$| is less than a chosen threshold. Our statistic is a credible interval, which is typically used for parameter estimation rather than hypothesis testing. However, credible intervals have been used for hypothesis tests against any threshold at the same credible level as the interval (*1–4*). The procedure checks if the threshold is within the bounds of the interval. In this sense, intervals can be used for hypothesis testing against any threshold without recalculation of the interval. We note that there is controversy with the use of credible intervals for hypothesis testing because the size of the effect is estimated under the assumption that it exists (*5*). We perceive this reservation to be a nonissue because an effect-size of zero has never been shown to exist in the real world and can't be confirmed with finite data (*6*). We assume a non-zero effect size is present in all cases, the question this procedure asks is whether there is evidence that it is small enough to be considered negligible.

We perform a hypothesis test against the specified threshold $\delta$ of the form

$$H_0: |r\mu_{DM}| \geq \delta; \quad H_1: |r\mu_{DM}| < \delta. \tag{S11}$$

We reject $H_0$ and conclude negligible effect size if r$\delta_M < \delta$ because r$\delta_M$ is the upper bound of the posterior for |r$\mu_{DM}$|.

**Contra Plot Axis Transform**

For this study, we produce contra plots that visualize the relative difference in means. This variable represents a percent change, which is difficult to visualize since a positive change is bound between [0, ∞] while a negative change is bound between [-1, 0]. This makes it impractical to visually compare effect sizes with difference signs since the distance from zero changes based on the sign of the effect. We could visualize this data in a log space to make the distance between the axis and each datapoint equal for positive and negative effect sizes, but we lose the ability to linearly compare effect sizes of the same sign in log space. To solve this issue, we used a transform we previously developed [ref] that distorts the negative axis so that negative signed effect sizes are the same distance to zero as their positive signed effect size while keeping the data in linear space. This transform allows scientists to compare positive and negative signed effect sizes, along with effect sizes that have the same sign. If X represents the input data (in this case the bounds of the interval estimates of relative effect size), we can transform the data with $f_{RC}(x)$, where

$$f_{RC}(x) = \begin{cases} x & if\ x \leq 0 \\ -\frac{1}{x+1} - 1 & otherwise \end{cases}. \tag{S12}$$

This transform encodes the position of each value in the cartesian space, but the axis markers are labeled with the data pre-transform. Please see publication for details [ref].

**Study Identification**

The studies presented in in Fig. 3-4 and STable 1-2 were compiled based on a literature search using Pubmed, Google Scholar, and Google search from our previous work on contra plots (*11*). Included results were limited to papers that were indexed on Pubmed. Papers were identified based on searches with combinations of the following keywords.

Total cholesterol example: atherosclerosis, total cholesterol, cholesterol, plasma cholesterol, reduce, protect, increase, independent, no change, mouse, rabbit, human, primate, rat.

Plaque size example: plaque size, plaque area, lesion size, lesion area, reduce, protect, increase, independent, no change, mouse, rabbit, human, primate, rat.

The included results are not meant to be complete, but rather give the reader a simplified toy example with how the proposed metric could be used to ascertain the practical insignificance of results.

The mean and standard deviation of each group were either copied directly from the source publication or estimated from the figure using Web Plot Digitizer (https://automeris.io/WebPlotDigitizer/)

# Supplementary Tables

Supplementary Table 1: Interventions tested for changes to total plasma cholesterol.

| ID | Study | Year | Group X | $\bar{x}$ | $s_x$ | $n_x$ | Group Y | $\bar{y}$ | $s_y$ | $n_y$ | Units | $\alpha_{DM}$ | Sp | PMID | Loc | Sgn |
|---|---|---|---|---|---|---|---|---|---|---|---|---|---|---|---|---|
| 1 | Billion | 2014 | Alfp-cre | 3.45 | 0.24 | 6 | Alfp-creTR?fl/fl | 3.26 | 0.22 | 6 | mmol/L | 0.05 | ms | 24797634 | F1C | 0 |
| 2 | Han | 2018 | ApoE-/-CD6WT | 1251 | 161 | 10 | ApoE-/-CD6-/- | 1179 | 143 | 5 | mg/dl | 0.05 | ms | 29615096 | T1 | 0 |
| 3 | Aherra. | 2020 | Cyp17AWT (F) | 2.29 | 0.53 | 7 | Cyp17A-/- (F) | 2.3 | 0.32 | 6 | mmol/L | 0.05 | ms | 32472014 | T2 | 0 |
| 4 | Yagyu | 2002 | VLDLr0-/- LpL1WT | 141 | 34 | 8 | VLDLr0-/- LpL1-/- | 151 | 24 | 13 | mg/dl | 0.05 | ms | 11790777 | T1 | 0 |
| 5 | Kirk | 1998 | LDLr-/- | 105 | 28 | 15 | LDLr-/- + IFD | 93.4 | 29 | 15 | mmol/L | 0.05/3 | ms | 9614153 | T2 | 0 |
| 6 | Han | 2018 | ApoE-/-CD6WT | 2518 | 257 | 8 | ApoE-/-CD6-/- | 2876 | 506 | 6 | mg/dl | 0.05 | ms | 29615096 | T1 | 0 |
| 7 | Niimi | 2013 | Vehicle | 1335 | 269 | 8 | Probucol | 1303 | 376 | 8 | mg/dL | 0.05/12 | rb | 24188322 | F1 | 0 |
| 8 | Cimen | 2016 | ApoE-/- | 568 | 81 | 6 | ApoE-/- PAO | 742 | 256 | 4 | mg/dL | 0.05 | ms | 27683551 | F5E | 0 |
| 9 | asahar. | 1994 | Control | 11.2 | 8.78 | 8 | Probucol | 15.3 | 6.55 | 8 | mg/dL | 0.05 | mc | 8040256 | T1 | 0 |
| 10 | Sjoland | 2000 | ApoE-/- PAI-1WT | 2503 | 266 | 11 | ApoE-/- PAI-1-/- | 1984 | 252 | 13 | mg/dl | 0.05 | ms | 10712412 | T1 | -1 |
| 11 | Ding | 2016 | Ldlr-/-Ad-Gal-RelnFl | 2087 | 531 | 15 | Ldlr-/- Ad-Cre-RelnF | 1487 | 364 | 16 | mg/dl | 0.05 | ms | 26980442 | SF2B | -1 |
| 12 | Niimi | 2013 | Vehicle | 1335 | 269 | 8 | Atorvastin | 934 | 231.9 | 8 | mg/dL | 0.05/12 | rb | 24188322 | F1 | -1 |
| 13 | Kirk | 1998 | WTD -IF | 4.78 | 1.2 | 20 | WTD +IF | 3.3 | 0.67 | 20 | mmol/L | 0.05/4 | ms | 9614153 | F1A | -1 |
| 14 | Milazzo | 2012 | Placebo | 202 | 28.2 | 5 | 150 mg Atorvastatin | 135.2 | 64.2 | 5 | mg/dL | 0.05/6 | hu | 22716983 | ST9 | -1 |
| 15 | Wu | 2020 | Luciferase siSRNA | 238 | 15.7 | 5 | Angptl3 siRNA | 153 | 22.4 | 5 | mg/dL | 0.05/6 | ms | 32808882 | F1D | -1 |
| 16 | Schuste | 2019 | WTD +Saline | 463 | 103 | 12 | WTD +PCSK9-mAb1 | 254 | 108 | 10 | mg/dl | 0.05 | ms | 31366894 | F1A | -1 |
| 17 | Rinaldi | 2000 | pCMV5 | 642 | 63 | 9 | PCMV-E3 | 283 | 69 | 10 | mg/dl | 0.05 | ms | 11110410 | F3 | -1 |
| 18 | Black | 2000 | ApoE-/- | 300 | 97 | 13 | ApoE-/- +Palm-E | 126 | 41 | 14 | mg/dl | 0.05 | ms | 11015467 | T1 | -1 |
| 19 | Lu | 2017 | ApoE-/- | 969 | 219 | 6 | ApoE–/– EC-TFEB | 965 | 84.6 | 6 | mg/dL | 0.05 | ms | 28143903 | SF6A | 0 |
| 20 | Zimmer | 2016 | ApoE-/- | 956 | 181 | 5 | ApoE–/– + CD | 770 | 61.3 | 4 | mg/dL | 0.05 | ms | 27053774 | F2F | 0 |
| 21 | Calleja | 1999 | ApoE-/- | 444 | 109 | 9 | ApoE–/– + Palm Oil | 324 | 88 | 9 | mg/dL | 0.05 | ms | 10521366 | T2 | 0 |
| 22 | Calleja | 1999 | ApoE-/- | 585 | 131 | 11 | ApoE–/– + SF Oil | 598 | 102 | 11 | mg/dL | 0.05/7 | ms | 10521366 | T3 | 0 |
| 23 | Han | 2018 | WTD-KO-F | 2338 | 332 | 5 | WTD-DKO-F | 2040 | 474 | 10 | mg/dL | 0.05 | ms | 29615096 | T2 | 0 |
| 24 | Zhuang | 2019 | Foxp1WTApoE-/- | 1037 | 635 | 12 | Foxp1ECKOApoE–/– | 932.5 | 442 | 12 | mg/dL | 0.05 | ms | 31318658 | ST1 | 0 |
| 25 | Kirk | 1998 | LDLr-/- WTD -IF | 105 | 27.9 | 15 | LDLr–/– WTD +IF | 93.4 | 29.4 | 15 | mmol/L | 0.05/3 | ms | 9614153 | T2 | 0 |
| 26 | Zhang | 2006 | Ldlr-/- FxrWT (M) | 1888 | 627 | 13 | Ldlr–/– Fxr–/– (M) | 1988 | 462 | 13 | mg/dL | 0.05/2 | ms | 16825595 | F3B | 0 |
| 27 | Yagyu | 2002 | LpL1WT | 104 | 16 | 8 | LpL1–/– | 146 | 34 | 9 | mg/dl | 0.05 | ms | 11790777 | T1 | 0 |
| 28 | Yagyu | 2002 | VLDlr0WT | 104 | 16 | 8 | VLDlr0–/– | 141 | 34 | 8 | mg/dl | 0.05 | ms | 11790777 | T1 | 0 |
| 29 | Moreira | 2012 | C57Bl/6 | 55.6 | 15 | 12 | LDLR-/- | 195.1 | 15.6 | 9 | mg/dL | 0.05/2 | ms | 22810096 | F1A | 1 |
| 30 | Zhang | 1992 | C57Bl/6 | 86 | 20 | 46 | ApoE-/- | 434 | 129 | 40 | mg/dL | 0.05/3 | ms | 1411543 | T1 | 1 |
| 31 | Zhang | 1992 | C57Bl/6 | 86 | 20 | 46 | ApoE+/- | 88 | 22 | 47 | mg/dL | 0.05/3 | ms | 1411543 | T1 | 0 |
| 32 | Ishib. | 1993 | C57Bl/6 | 119 | 17 | 19 | LDLR-/- | 228 | 36 | 16 | mg/dL | 0.05/3 | ms | 8349823 | T1 | 1 |
| 33 | Ishib.i | 1993 | C57Bl/6 | 119 | 17 | 19 | LDLR+/- | 158 | 25 | 39 | mg/dL | 0.05/3 | ms | 8349823 | T1 | 0 |
| 34 | Hartv. | 2006 | LDLR-/- | 8.7 | 2.6 | 12 | LDLR-/- HC | 21.45 | 5.2 | 12 | mmol/L | 0.05/3 | ms | 17255537 | F2A | 1 |
| 35 | Hartv. | 2006 | LDLR-/- | 8.7 | 2.6 | 12 | LDLR-/- WD | 48.88 | 7.6 | 8 | mmol/L | 0.05/3 | ms | 17255537 | F2A | 1 |

*Abbreviations: ID, study identification number to compare studies between contra-plot and this supplemental table; Group X, control group label; $\bar{x}$, control group sample mean; $s_x$, control group sample standard deviation; $n_x$, control group sample size; group y, experiment group label; $\bar{y}$, experiment*

*group sample mean; $s_y$, experiment group sample standard deviation; $n_y$, experiment group sample size; Units, units of measure for variable of interest; $a_{DM}$, Bonferroni correction required for interval estimate; Sp, ; PMID, pubmed identification number of study; Loc, figure or table location of data in manuscript; Sgn, sign/ direction of effect size as reported from study (0 if result was reported as statistically insignificant).*

Supplementary Table 2: Interventions tested for changes to change plaque size

| ID | Study | Year | Group X | $\bar{x}$ | $s_x$ | $n_x$ | Group Y | $\bar{y}$ | $s_y$ | $n_y$ | Units | $\alpha_{DM}$ | Sp | PMID | Loc | Sgn |
|---|---|---|---|---|---|---|---|---|---|---|---|---|---|---|---|---|
| 1 | Ulrich | 2016 | Saline | 48.9 | 3.3 | 9 | LNA-Control | 51.6 | 5.2 | 13 | % | 0.05/3 | ms | 27137489 | F1C | 0 |
| 2 | Santiago. | 2020 | ApoE-/- | 6.7E+05 | 1.1E+05 | 8 | ApoE-/-.Yaa | 6.7E+05 | 1.6E+05 | 8 | µm2 | 0.05 | ms | 33110193 | F1A | 0 |
| 3 | Flach | 2019 | Vehicle | 3.7E+05 | 1.3E+05 | 14 | PF-'2999 | 3.6E+05 | 1.6E+05 | 15 | µm2 | 0.05/1 | ms | 30889221 | F2B | 0 |
| 4 | Kirk | 1998 | LDLr-/- -IF | 360 | 94 | 9 | LDLr-/- +IF | 295 | 36 | 8 | mm2 | 0.05/1 | ms | 9614153 | F2A | 0 |
| 5 | Zhang | 2006 | LDLr-/- FxrWT | 23.1 | 11.3 | 7 | LDLr-/- Fxr-/- | 19.4 | 7.2 | 8 | % | 0.05/2 | ms | 16825595 | F2B | 0 |
| 6 | Lyon | 1987 | Prox. -Sten. | 36 | 23 | 15 | Prox. +Sten. | 41 | 50 | 13 | % | 0.05 | mk | 3795393 | T3 | 0 |
| 7 | Bocan | 1994 | Vehicle | 0.713 | 0.297 | 8 | BMY22089 | 0.617 | 0.45 | 8 | mm2 | 0.05/7 | rb | 7840808 | T4 | 0 |
| 8 | Nichols | 1999 | Vehicle | 304788 | 113425 | 4 | SC-64258 | 310284 | 160647 | 3 | µm2 | 0.05/3 | pg | 10571535 | T2 | 0 |
| 9 | Wang | 2011 | ApoE-/- -Abx | 9956 | 11578 | 20 | ApoE-/- +Abx | 17196 | 13373 | 18 | µm2 | 0.05/6 | ms | 21475195 | F5E | 0 |
| 10 | Hechler | 2008 | ApoE-/- | 0.54 | 0.12 | 16 | ApoE-/-P2Y1-/- | 0.41 | 0.1162 | 15 | mm2 | 0.05 | ms | 18663083 | F2B | -1 |
| 11 | Cimen | 2016 | ApoE-/- | 225000 | 72732 | 10 | ApoE-/- + PAO | 161000 | 47434 | 10 | µm2 | 0.05 | ms | 27683551 | F5C | -1 |
| 12 | Hechler | 2008 | ApoE-/- | 21.2 | 8.08332 | 6 | ApoE-/-P2Y1-/- | 12.4 | 2.9394 | 6 | % | 0.05 | ms | 18663083 | F1A | -1 |
| 13 | Lu | 2017 | ApoE-/- | 16.1 | 7.7 | 10 | ApoE-/-EC-TFEB | 8.58 | 3.3 | 12 | % | 0.05 | ms | 28143903 | F7F | -1 |
| 14 | Nichols | 1999 | Vehicle | 304788 | 113425 | 4 | SC-69000 | 149779 | 34576 | 7 | µm2 | 0.05/3 | pg | 10571535 | T2 | -1 |
| 15 | Fang | 2018 | ApoE-/- | 46.2 | 10.6 | 3 | ApoEWT | 22.1 | 8 | 3 | % | 0.05 | pg | 30305304 | F5A | -1 |
| 16 | Bocan | 1994 | Progression | 0.713 | 0.297 | 8 | Atorvastatin | 0.221 | 0.147 | 8 | mm2 | 0.05/7 | rb | 7840808 | T4 | -1 |
| 17 | Kirk | 1998 | LDLr-/- -IF | 7.39 | 6.7 | 13 | LDLr-/- +IF | 2.01 | 3.4 | 15 | % | 0.05 | ms | 9614153 | F2B | -1 |
| 18 | Niimi | 2013 | Vehicle | 0.173 | 0.15 | 8 | Probucol | 0.015 | 0.0255 | 8 | mm2 | 0.05/6 | rb | 24188322 | F1 | -1 |
| 19 | Lv | 2017 | Control | 0.0386 | 0.04554 | 10 | LDLR-/- | 0.86 | 0.2814 | 10 | mm2 | 0.05/10 | ms | 28983592 | F1D | 1 |
| 20 | Ma | 2012 | Chow | 11904 | 8874.14 | 4 | HFD | 859127 | 210628 | 4 | um2 | 0.05/15 | ms | 22558236 | F2B | 1 |
| 21 | Wang | 2021 | LDLR-/- | 461.5 | 131.9 | 9 | LDLR-/-FbnC1039G+/- | 747.3 | 153.8 | 9 | um2 | 0.05/4 | ms | 33796572 | F4A | 1 |
| 22 | Zhang | 2020 | LDLR-/- | 0.091 | 0.041 | 3 | PM2.5 | 0.219 | 0.056 | 3 | mm2 | 0.05/4 | ms | 31935561 | F2E | 1 |
| 23 | Gao | 2010 | Isotype | 189413 | 99532.7 | 10 | a-IL-17A | 10237 | 35547 | 10 | um2 | 0.05 | ms | 20952673 | F5A | -1 |
| 24 | Engel. | 2019 | WT | 2.2E+05 | 6.5E+04 | 17 | CD47-/- | 3.4E+05 | 1.9E+05 | 17 | um2 | 0 | ms | 31337788 | F1C | 1 |
| 25 | Boriss. | 2013 | ApoE-/-FII-/+ | 9.1E+05 | 3.5E+05 | 10 | ApoE-/- | 1.6E+06 | 6.6E+05 | 10 | um2 | 0.05 | ms | 23409043 | F1B | 1 |
| 26 | Brampt. | 2021 | LDLR-/- | 20.13 | 2.42 | 6 | LDLR-/-Abcc6-/- | 39.3 | 5.22 | 7 | % | 0.05/3 | ms | 33594095 | F1A | 1 |
| 27 | Park | 2016 | NFD | 0.296 | 0.11 | 5 | HFD | 8.33 | 1.52 | 3 | mm2 | 0.05/6 | ms | 26950217 | F1B | 1 |
| 28 | Park | 2016 | LDLR-/- | 640.7 | 175.7 | 10 | LDLR-/-Creb3l3-/- | 854.4 | 157.5 | 10 | um2 | 0.05 | ms | 27417587 | F2C | 1 |

*Abbreviations: see Table S1.*

## Supplementary References